\documentclass[11pt]{article} 
\usepackage{amsfonts,amscd,amsmath,graphicx}
\newcommand{\oh}{\frac{1}{2}}
\def\4{\tfrac{1}{4}}
\def\ep{\text{e}}

\def\g{\mathfrak{g}}
\def\h{h_{\text{\tiny 0}}}
\def\z{z_{\text{\tiny 0}}}

\title{Heavy-Quark  Potentials and AdS/QCD}
\author{Oleg Andreev\thanks{andre@itp.ac.ru}\\\\
{\it L.D. Landau Institute for Theoretical Physics, Kosygina 2, 119334 Moscow, Russia}\\\\
Valentin I. Zakharov\thanks{xxz@mppmu.mpg.de}\\\\
{\it Istituto Nazionale di Fisica Nucleare -- Sezione di Pisa} \\
{\it Dipartimento di Fisica Universita di Pisa, Largo Pontecorvo 3, 56127 Pisa, Italy;}\\
{\it Max-Planck Institut f\" ur Physik, F\" ohringer Ring 6, 80805 M\" unchen, Germany} }
\date{}

\begin{document} 

\vspace{-8cm} 
\maketitle 
\begin{abstract} 
We give an example of modeling phenomenological heavy-quark potentials in a five-dimensional framework nowadays known 
as AdS/QCD. In particular we emphasize the absence of infrared renormalons.
 \end{abstract}

\vspace{-13cm}
\begin{flushright}
LANDAU-06-A1
\end{flushright}
\vspace{12.5cm}

\section{Introduction}

Heavy-quark potential is one of the basic observables relevant to confinement.
It has been measured in great detail in lattice simulations\footnote{For a review, see \cite{bali}.}
and the results reveal a remarkable  agreement with the so called Cornell potential \cite{cornell}

\begin{equation}\label{cornell}
V(r)=-\frac{\kappa}{r}+\frac{r}{a^2}+C
\,,
\end{equation}
where the coefficients are adjusted to fit the charmonium spectrum

\begin{equation}\label{coeff}
\kappa\approx 0.48\,,
\quad\quad
a\approx 2.34\,\,\text{GeV}^{-1}
\,,
\quad\quad
C=-0.25\,\,\text{GeV}
\,.
\end{equation}
As follows from above, $\kappa$ and $\tfrac{1}{a^2}$ can be interpreted as $\tfrac{4}{3}\alpha_s$ and 
the string tension, respectively. 

One of the implications of the AdS/CFT correspondence \cite{ads/cft} is that it resumed interest in finding a string description of 
strong interactions. For the case of interest, let us briefly mention a couple of results. 

First, in the approach called usually gauge/string duality one tries to keep the underlying string structure. As a consequence, the 
theory is ten dimensional and its reductions to five dimensions in general contain additional higher derivative 
terms (stringy $\alpha'$ corrections).\footnote{These corrections are of order $\tfrac{1}{\sqrt{N_c}}$. Thus, they might be relevant 
at $N_c=3$.} According to \cite{malda}, the expectation value of the Wilson loop is given by 

\begin{equation}\label{wloop}
\langle\,W({\cal C})\,\rangle \sim \ep^{-S}
\,,
\end{equation}
where $S$ is an area of a string world-sheet bounded by a curve $\cal C$ at the boundary of AdS space.\footnote{The literature on the 
Wilson loops within the AdS/CFT correspondence is very vast. For a discussion of this issue, see, e.g., \cite{loops} and references therein.}

Second, in more phenomenological approach called AdS/QCD one starts from a five dimensional effective field theory somehow 
motivated by string theory and tries to fit it to QCD as much as possible. It was recently pointed out \cite{katz, oa} that asymptotic 
linearity of Regge trajectories arises for some backgrounds. Such backgrounds reduce to the standard AdS background in the UV  but 
differ from it in the IR. The latter turns out to be crucial for linearity. In this case, it would be natural to expect that the interquark 
interaction would include the dominant Coulomb term at short distances as well as the dominant linear term at large distances.

In this note we explore this expectation in the context of \eqref{wloop}. For simplicity we will concentrate here on the 
case of \cite{oa}. However, it should not be hard to adapt the arguments to \cite{katz}. It is worth noting a recent attempt to 
derive the Cornell type potential within the model based on a truncated AdS space \cite{cornell-ads}.\footnote{In this case there is a 
subtle point. The use of two different solutions leads to a discontinuity in the interquark force.}

Before proceeding to the detailed analysis, let us set the basic framework. We consider the following Euclidean background metric

\begin{equation}\label{metric}
ds^2=G_{nm}dX^ndX^m=
R^2\frac{h}{z^2}
\left(dx^i dx^i+dz^2\right)
\,,\quad\quad h=\ep^{\oh cz^2}\,,
\end{equation}
where $i=0,\dots ,3$. In the region of small $z$ the metric behaves asymptotically as Euclidean $\text{AdS}_5$, as expected. We  also 
take a constant dilaton. Note that the use of the Euclidean signature for the background metric slightly modifies $h$. So, the exponent 
has an opposite sign to that of \cite{oa}.
 

\section{Calculating the Potential}

Given the background metric,  we can attempt to calculate the corresponding potential. In doing so, we adapt the 
conjecture \eqref{wloop} to AdS/QCD. 

We consider a rectangular Wilson loop $\cal{C}$ living on the boundary ($z=0$) of five dimensional space as shown in Fig.1. 
%
\begin{figure}[ht]
\begin{center}
\includegraphics{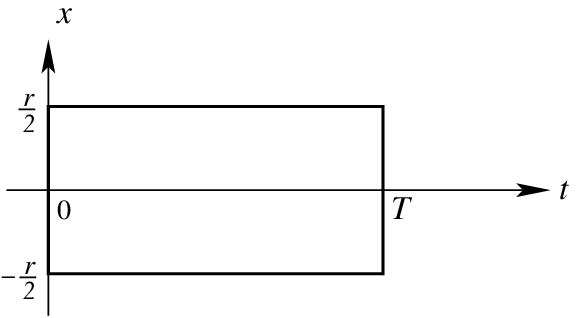}
\caption{\small{A loop $\cal C$.}}
\end{center}
\end{figure}
The quark and antiquark are set at $x=\tfrac{r}{2}$ and $x=-\tfrac{r}{2}$, respectively. As known, taking the 
limit $T\rightarrow\infty$ allows one to read off the energy of such a pair from the expectation value of the Wilson loop namely, 
$\langle W({\cal C})\rangle \sim \ep^{-TE(r)}$.

Now we are ready to evaluate the expectation value of the loop. To this end, we make use of the Nambu-Goto action equip with the 
background metric \eqref{metric}

\begin{equation}\label{ng}
S=\frac{1}{2\pi\alpha'}\int d^2\xi\,\sqrt{\det \, G_{nm}^{}\partial_\alpha X^n\partial_\beta X^m\vphantom{\bigl(\bigr)}}
\,.
\end{equation}
Next, we choose $\xi^1=t$ and $\xi^2=x$. This yields

\begin{equation}\label{ng1}
S=\frac{\g }{2\pi}T\int^{\,\tfrac{r}{2}}_{-\tfrac{r}{2}} 
dx\,\frac{h}{z^2}
\sqrt{1+(z')^2}
\,,
\end{equation}
where $\g=\tfrac{R^2}{\alpha'}$. A prime denotes a derivative with respect to $x$.

Now it is easy to find the equation of motion for $z$
\begin{equation}\label{eqm}
zz''+\left(2-cz^2\right)\left(1+(z')^2\right)=0
\end{equation}
as well as the first integral
\begin{equation}\label{int}
\frac{h}{z^2\sqrt{1+(z')^2}}=\text{const}
\,.
\end{equation}
The integration constant can be expressed via the maximum value of $z$. On symmetry grounds, $z$ reaches it at $x=0$. By 
virtue of \eqref{int}, the integral over $\left[-\tfrac{r}{2},\tfrac{r}{2}\right]$ of $dx$ is equal to 

\begin{equation}\label{z0}
r=2 \sqrt{\frac{\lambda}{c}}\int^1_0 dv\,v^2\ep^{\tfrac{1}{2}\lambda (1-v^2)}
\Bigl(1-v^4\ep^{\lambda (1-v^2)}\Bigr)^{-\tfrac{1}{2}}
\,,
\end{equation}
where $v=\tfrac{z}{z_0}$, $\lambda=cz_0^2$, and $z_0=z\vert_{x=0}$.

At this point a comment is in order. A simple analysis shows that the integral \eqref{z0} is real for $\lambda <2$. It develops a 
logarithmic singularity at $\lambda=2$ and becomes complex for larger $\lambda$. Hence, there exists the upper bound on the 
maximum value of $z$
\begin{equation}\label{upper}
z_0<\sqrt{\frac{2}{c}}
\,.
\end{equation}
Note that in the limit as $c$ goes to zero $z_0$ is not bounded, as should be for the $\text{AdS}$ space.
As function of $z$, the effective string tension reaches its minimum at $z=z_0$. 
Thus, there exists a kind of horizon which is a generic feature of confining theories. 
 
Now, as in \cite{malda}, we will compute the energy of the configuration. 
First, we reduce the integral over $x$ in Eq.\eqref{ng1} to that 
over $z$. This is done by using the first integral \eqref{int}. Since the integral is divergent at $z=0$ due to the factor $z^{-2}$ in the 
metric \eqref{metric}, in the process we regularize it by imposing a cutoff $\epsilon$. Then we replace $z$ with $v$ as in \eqref{z0}. 
Finally, the regularized expression takes the form

\begin{equation}\label{energy}
E_{\text{\tiny R}}=\frac{\g}{\pi}\sqrt{\frac{c}{\lambda}}\int^1_{\tfrac{\epsilon}{z_0}}dv\,v^{-2}
\ep^{\tfrac{1}{2}\lambda v^2}
\Bigl(1-v^4\ep^{\lambda (1-v^2)}\Bigr)^{-\tfrac{1}{2}}
\,.
\end{equation}
Its $\epsilon$-expansion is simply 

\begin{equation}\label{energy1}
E_{\text{\tiny R}}=\frac{\g}{\pi\epsilon}+E+O(\epsilon)
\,,
\end{equation}
where 

\begin{equation}\label{energy-f}
E=\frac{\g}{\pi}\sqrt{\frac{c}{\lambda}}
\left(-1+
\int^1_0dv\,v^{-2}\Bigl[
\ep^{\tfrac{1}{2}\lambda v^2}
\Bigl(1-v^4\ep^{\lambda (1-v^2)}\Bigr)^{-\tfrac{1}{2}}-1
\Bigr]
\right)
\,.
\end{equation}

Similarly as $r$, $E$ is real only for $\lambda <2$. Having observed that the energy acquires an imaginary part, it is tempting to interpret 
this as the string breaking. However, this occurs at complex $r$. Our model is therefore stable.

In contrast to the $\text{AdS}$ case \cite{malda},  the potential in question is written in parametric form given by Eqs.\eqref{z0} and 
\eqref{energy-f}.\footnote{It is unclear to us how to eliminate the parameter $\lambda$ and find $E$ as a function of $r$ and $c$.} We 
can, however, gain some important insights into the problem by considering two limiting cases. 

First, let us have a close look at Eq.\eqref{z0}. As noted earlier, the range of $\lambda$ is $0\leq\lambda <2$. After a short inspection 
we find that $r$ is a continuously growing function of $\lambda$. The  asymptotic behavior near zero is given by \footnote{All the 
integrals can be evaluated in terms of the beta functions as discussed in the appendix A.}

\begin{equation}\label{r0}
r=\frac{1}{\rho }\sqrt{\frac{\lambda}{c}}\,
\biggl(
1-\frac{1}{4}\lambda\bigl(1-\pi\rho^2\bigr)+O(\lambda^2)
\biggr)
\,,
\end{equation}
where $\rho=\Gamma^2\left(\tfrac{1}{4}\right)/(2\pi)^{\tfrac{3}{2}}$. From this it follows that small $\lambda$'s correspond to 
small values of $r$. 

The asymptotic behavior near $2$ is given by \footnote{The integral is dominated by $v\sim 1$, where it takes the following form 
$2\sqrt{\tfrac{\lambda}{c}}\int_0^1 dv /\sqrt{a(1-v)+b(1-v)^2}$, with $a=2(2-\lambda)$ and 
$b=\lambda-(2-\lambda)(3+2\lambda)$. The remaining integral may be found in tables \cite{gr}.}

\begin{equation}\label{r2}
r=-\sqrt{\frac{2}{c}}\ln (2-\lambda) +O(1)
\,.
\end{equation}
Thus, this region corresponds to large values of $r$.

Having understood the correspondence between $\lambda$ and $r$, we can investigate the properties of the interquark interaction at 
short and long distances. 

We begin with the case of small $r$. Expanding the right hand side of Eq.\eqref{energy-f} up to the quadratic terms in $\lambda$, 
we get

\begin{equation}\label{E}
E=-\frac{\g}{2\pi\rho}\sqrt{\frac{c}{\lambda}}
\biggl(
1+\frac{1}{4}\lambda\bigl(1-3\pi\rho^2\bigr)
+O(\lambda^2)
\biggr)
\,.
\end{equation}
Combining this with \eqref{r0}, we find the energy of the configuration as a function of $r$ and $c$

\begin{equation}\label{small}
E=\g\Bigl(-\frac{\kappa_{\text{\tiny 0}} }{r}+\sigma_{\text{\tiny 0}} r+O\bigl(r^3\bigr)\Bigr)
\,,
\end{equation}
where $\kappa_{\text{\tiny 0}} =\tfrac{1}{2\pi}\rho^{-2}$ and $\sigma_{\text{\tiny 0}}=\tfrac{1}{4}c\rho^2$. Thus, we have the 
expected $1/r$ behavior  at short distances.

In a similar spirit, we can explore the long distance behavior of $E$. It follows from \eqref{energy-f} that in the neighbor 
of $\lambda=2$ the energy behaves as  

\begin{equation}\label{E2}
E=-\frac{\ep\g}{2\pi}
\sqrt{\frac{c}{2}}\ln (2-\lambda) +O(1)
\,.
\end{equation}
Along with the relation \eqref{r2}, this means that the interquark interaction at long distances is given by

\begin{equation}\label{E2-1}
E=\g\Bigl(\sigma r+O(1)
\Bigr)
\,
\end{equation}
that is nothing but the desired linear potential. Here we have set  $\sigma=\tfrac{\ep}{4\pi} c$.

Having understood the two limiting cases, we can now make a couple of estimates
relevant to phenomenology.

It is natural to fix the overall constant $\g$ from the slope of the potential at large
distances. Indeed, the stringy approach is to be most reliable at large distances.
From \eqref{cornell} and \eqref{E2-1} we have 

\begin{equation}\label{g}
\g=\frac{4\pi}{\ep} \bigl(c a^2 \bigr)^{-1} \approx 0.94
\,,
\end{equation}
where we have used $c\approx 0.9\,\text{GeV}^2$ as it follows from the fits to the slope of the Regge trajectories \cite{oa}.

Next, we can estimate the slope of the linear potential at short distances. According to
the Cornell model (\ref{cornell}), the slope is the same at all the distances
while Eqs.(\ref{E2-1}) and (\ref{small}) imply that the coefficients in front of the linear
terms
at large and small distances are different. 
However, a simple estimate 
of their ratio yields 

\begin{equation}\label{ratio}
\frac{\sigma}{\sigma_{\text{\tiny 0}}}=8\pi^2\ep\,\Gamma^{-4}(\tfrac{1}{4})\approx 1.24
\,.
\end{equation}
Clearly, the difference in the slopes is not significant for our phenomenological estimates
and the agreement with the lattice data is very satisfactory at this point. 

Finally, we can compare evaluate the $1/r$ term in the potential. Phenomenologically,
a little algebra shows that in \eqref{cornell} the coefficients obey $1/\kappa a^2\approx 0.38\,\text{GeV}^2$. 
If we truncate our 
model by keeping only the two terms as in the Cornell model, we find

\begin{equation}\label{Cl}
\frac{\sigma}{\kappa_{\text{\tiny 0}}}=\frac{1}{16\pi^3}\,\ep\,\Gamma^4(\tfrac{1}{4})\,c
\approx 0.85\,\text{GeV}^2
\,.
\end{equation}
The value is more than twice bigger than that of the Cornell model. So, this looks disappointing. However, the Coulomb-like 
potential at short distances is controlled by the running coupling $\alpha_s(r)$ and can hardly be predicted within
the simplified stringy model we are considering. We will come back to discuss this point 
in the next section.

To complete the picture, let us present the results of numerical calculations. The parametric equation \eqref{z0} predicts a 
characteristic form for $r$, as shown in Fig.2.
%
\begin{figure}[ht]
\begin{center}
\includegraphics[width=7cm]{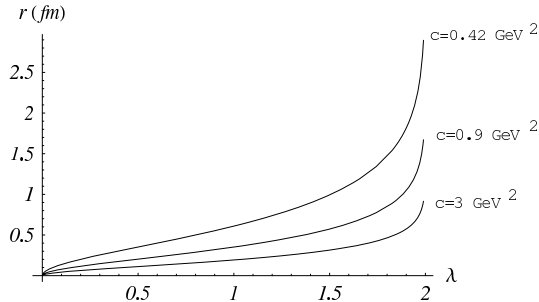}
\caption{\small{$r$ as a function of $\lambda$ at $c=0.42,\,\,0.9,\,\,\text{and}\,\,3\,\,\text{GeV}^2$.}}
\end{center}
\end{figure}
It has an interesting effect on the form of the interquark potential in the phenomenologically important interval $0.1\,\text{fm}\,
\leq r\leq 1\,\text{fm}$.  It is clear that for quite small values of $c$ this interval corresponds to small $\lambda$'s. As a result, 
the approximate formula \eqref{small} is valid. So, if the desired linear behavior holds, it has the slope proportional to 
$\sigma_{\text{\tiny 0}}$. On the other hand,  larger values of $c$ result in $\lambda'\text{s}\sim 2$. In this case the linear 
behavior has the slope proportional to $\sigma$. The effect can be seen in Fig.3 for the window 
$0.5\,\text{fm}\,\leq r\leq 1\,\text{fm}$ and the two values of $c$  namely, $c=0.42\,\,\text{and}\,\,3\,\text{GeV}^2$.
%
\begin{figure}[ht]
\begin{center}
\includegraphics[width=7cm]{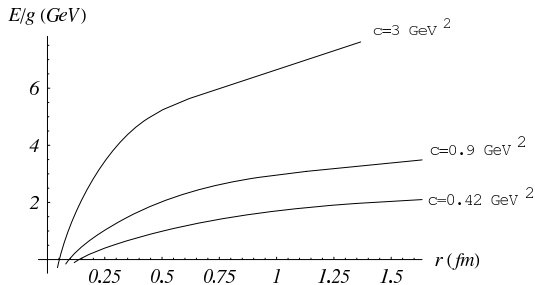}
\caption{\small{$E/\g$ as a function of $r$ at $c=0.42,\,\,0.9,\,\,\text{and}\,\,3\,\,\text{GeV}^2$.}}
\end{center}
\end{figure}

\section{Discussions}

In this note we  have evaluated the heavy-quark potential using the now standard ideas 
motivated by the gauge/string duality. A key point is the use of the background metric
\eqref{metric} which is singled out by the observation \cite{oa} that it provides,
in terms of the same duality, linear Regge trajectories. 
The overall conclusion is that the same background metric results in a phenomenologically satisfactory
description of the confining potential as well. However, there is a number of open problems
which are mostly rooted in the heuristic nature of the gauge/string duality in the case 
of pure Yang-Mills theory. In conclusion, we list a few such problems and compare the results 
with theoretical expectations based on more traditional approaches.

(i) The potential of interest behaves as $1/r$ at short distances.  This behavior is  
predicted, of course, by perturbative QCD as well. On the string theory side, one might fix the overall factor $\g$ in the 
world-sheet action \eqref{ng1} by fitting the coefficients of the $1/r$ terms in the energy \eqref{E} 
and the perturbative calculations. If so, then $\g\sim\alpha_s$. By contrast, the AdS/CFT correspondence requires $\g\sim\sqrt{\alpha_s}$ 
\cite{ads/cft}. Thus, this way of fixing the overall factor does not look satisfactory.\footnote{This difficulty was pointed out in different 
contexts. See, e.g, \cite{malda, oa2}. } 

On the other hand, in the case of pure Yang-Mills theory the status of the $1/r$ term in the 
potential (\ref{cornell}) is also unclear. 
Its numerical value fits the theoretical prediction for the so called L\"{u}scher term 
\cite{luscher} which is derived at large
distances. Phenomenologically the fit (\ref{cornell}) works at small distances as well.
Moreover, a straightforward application of the expansion in the running coupling 
$\alpha_s(r)$ results in a badly divergent series at presently available distances $r$.\footnote{For a review, see \cite{bali}.} Thus, 
it is not ruled out that  a pure perturbative description 
sets in  at much smaller distances. In view of all these theoretical uncertainties, the lack of agreement 
between the stringy potential (\ref{small}) and the Cornell model at short distances  might be not
so significant.
 
(ii) On the lattice, the so called Casimir scaling has been observed \cite{bali}. The point is that if one measures the
heavy-quark potential for various color representations of the quarks, at large distances the string tension turns out to be 
proportional to the coefficient of the Coulomb-like term. Reproducing the Casimir scaling theoretically is a strong challenge to 
the model building \cite{simonov}. Within AdS/QCD, such a scaling is obvious. It is reproduced without fixing any parameter.
Indeed, the fitting parameter is the overall factor $\g$ in the expression for the energy, see the discussion above.

 (iii) Power-like corrections to the heavy-quark potential at short distances were studied earlier mostly in terms of infrared 
renormalons.\footnote{For further discussion, see \cite{beneke}.} Within this approach, the potential can be represented, in 
somewhat symbolical form, as 

\begin{equation}\label{renormalon}
V(r)=
(\text{perturbative series})+\sum_{k=0} b_k\,r^{2k}\Lambda_{\text{QCD}}^{2k+1} 
\,,
\end{equation}
where the coefficients $b_k$ cannot be determined consistently within short-distance physics and correspond to the infrared 
renormalons. In particular, the leading non-perturbative correction to the potential
is of order \cite{balitsky}

\begin{equation}\label{balitsky}
V_{\text{non-pert}}\sim r^2\langle\,\alpha_s G^2\,\rangle \rho_{\text{inst}}
\,,
\end{equation}
where $\langle\alpha_sG^2\rangle$ is the gluon condensate and $\rho_{\text{inst}}$ is a typical instanton size.
Note that the instanton size plays the role of an infrared cutoff. Without such a cutoff, the calculation is divergent.

In our example, the potential is calculable consistently at short distances. There are no uncertainties 
corresponding to the infrared renormalons. The only exception might be the leading renormalon corresponding to $k=0$ 
in \eqref{renormalon}. Indeed, by the potential one can understand the difference between the total energy 
and self-energy of the quarks. In the bulk of the paper, we were calculating the total energy.
Separation of the potential from the self-energy involves then the infrared uncertainty in the mass
of the heavy quark which corresponds to the renormalon.\footnote{At technical level this means that the use of minimal subtraction 
in Eq.\eqref{energy1} is not appropriate.} 

We close the discussion of the infrared renormalons with a few short comments:
\newline 1. There exists a simple picture of the absence of the renormalons. The key point is that at short distances 
a string doesn't go far away into the fifth dimension or, in other words, it doesn't reach a vicinity of the horizon that would 
correspond to the IR.
\newline 2. We discuss a class of metric leading to the absence of the renormalons with $k>0$  in the appendix B. 
\newline 3. Phenomenologically, there is no significant $r^2$ term in the potential.\footnote{For further discussion, see
\cite{sumino}.} Thus, the absence of the infrared renormalons from the stringy potential can be considered 
as a success of the model.

(iv) The AdS/QCD approach provides a natural framework for appearance of a linear piece in the potential at short distances.
Moreover, as follows from Eq.\eqref{ratio}, the coefficients of the linear terms in the short and large distance expansions turn out 
to be close to each other. Phenomenological arguments in favor of such a contribution can be found in \cite{narison}.

\vspace{.25cm}
{\bf Acknowledgments}

\vspace{.25cm}
O.A. would like to thank P. Weisz for useful discussions. The  work  of O.A. was supported in part by the Alexander von Humboldt 
Foundation and Russian Basic Research Foundation Grant 05-02-16486. O.A. would also like to acknowledge the hospitality of the 
Heisenberg Institut, where a main portion of this work was completed.


\vspace{.35cm} 
{\bf Appendix A}
\renewcommand{\theequation}{A.\arabic{equation}}
\setcounter{equation}{0}

\vspace{.25cm} 
\noindent This appendix collects together some of the formulae that are used in section 2.

All the integrals can be expressed in terms of integrals of the form

\begin{equation}\label{bint}
I(a,b)=\int_0^1 dv\,v^a \left(1-v^4\right)^b
\,.
\end{equation}
Making the variable change $w=v^4$, we can easily evaluate the integral over $w$. The final result is 

\begin{equation}\label{bint1}
I(a,b)=\tfrac{1}{4}B\left(\tfrac{a+1}{4}, b+1\right)
\,,
\end{equation}
where $B$ denotes the beta function. 

In manipulating the beta functions, we use the relation to the gamma function and the formulae \cite{gr}

\begin{equation}\label{ex}
\Gamma\left(\tfrac{1}{2}\right)=\sqrt{\pi}\,,
\quad\quad
\Gamma\left(\tfrac{3}{4}\right)\Gamma\left(\tfrac{1}{4}\right)=\sqrt{2}\pi
\,,
\quad\quad
\Gamma\left(x+1\right)=x\Gamma\left(x\right)
\,.
\end{equation}

\vspace{.35cm} 
{\bf Appendix B}
\renewcommand{\theequation}{B.\arabic{equation}}
\setcounter{equation}{0}

\vspace{.25cm} 
\noindent In this appendix we will investigate the question of whether for a generic  form of the warp factor $h(z)$ in the 
metric \eqref{metric} the renormalons with $k>0$ are missing as well.\footnote{We assume $h(z)>0$.}

Since at $z=0$ the metric reduces to that of $\text{AdS}_5$, we take 

\begin{equation}\label{h0}
h(0)=1
\,. 
\end{equation}

The Nambu-Goto action and the first integral of equation of motion are given by Eq.\eqref{ng1} and Eq.\eqref{int}, respectively. 
From the first integral we get 
\begin{equation}\label{rh}
r=2\z\int_0^1 dv\,v^2 \,\frac{\h}{h}\biggl(1-v^4\Bigl(\frac{\h}{h}\Bigr)^2\biggr)^{-\oh}
\,,
\end{equation}
where $\z$ is the maximum value of $z$, $v=\tfrac{z}{\z}$, and $\h=h(\z)$.

We compute the energy of the configuration as in section 3. In the process we regularize the integral over $z$ by 
imposing a cutoff $\epsilon$. Finally, we get

\begin{equation}\label{energyh}
E_{\text{\tiny R}}=\frac{\g}{\pi\epsilon}+E+O(\epsilon)
\,,
\end{equation}
where 

\begin{equation}\label{energyh1}
E=\frac{\g}{\pi\z}
\left(-1+
\int^1_0dv\,v^{-2}\Bigl[h
\Bigl(1-v^4\Bigl(\frac{\h}{h}\Bigr)^2\Bigr)^{-\tfrac{1}{2}}-1
\Bigr]
\right)
\,.
\end{equation}
As a result, the potential is written in parametric form given by Eqs.\eqref{rh} and \eqref{energyh1}. 

Assume that we eliminated the parameter $\z$ and found $E$ as a function of $r$.\footnote{ As in section 3, we should 
consider a range of $\z$ corresponding to real $r$ and $E$. For small $r$ (equivalently, small $\z$) this is indeed the case.} Now 
the question arises: when does this potential have no terms like $r^{2k}$? The answer to this is clear from the form of equations. If 
we transform $\z\rightarrow -\z$, then $r\rightarrow -r$ for any even function $h$. Meanwhile, the energy transforms as 
$E\rightarrow -E$. Thus, $E$ is an odd function of $r$ if $h$ is even. The latter means that in this case the problem of the 
renormalons with $k>0$ is missing.


\end{document}